\documentclass[12pt,preprint]{aastex}

\slugcomment{Accepted for publication in the Astrophysical Journal}
\shorttitle{White Dwarf Stars in M5}
\shortauthors{Layden \etal}

\begin{document}


\title{Deep Photometry of the Globular Cluster M5: \\
Distance Estimates from White Dwarf and Main Sequence Stars}


\author{Andrew C. Layden} 
\affil{Department of Physics and Astronomy, 104 Overman Hall, \\
Bowling Green State University, Bowling Green, OH 43403 } 
\email{layden@baade.bgsu.edu}

\author{Ata Sarajedini} 
\affil{Department of Astronomy, 211 Bryant Space Science Center, P.O. Box 
112055, \\
University of Florida, Gainesville, FL 32611}
\email{ata@astro.ufl.edu} 
 
\author{Ted von Hippel} 
\affil{Department of Astronomy, 1 University Station C1400, \\
The University of Texas at Austin, Austin, TX 78712-0259}
\email{ted@astro.as.utexas.edu} 

\and

\author{Adrienne M. Cool} 
\affil{Department of Physics and Astronomy, 1600 Holloway Avenue, \\
San Francisco State University, San Francisco, CA 94132}
\email{cool@sfsu.edu} 



\begin{abstract}

We present deep $VI$ photometry of stars in the globular cluster M5
(NGC~5904) based on images taken with the {\it Hubble Space
Telescope}.  The resulting color-magnitude diagram reaches below $V
\approx 27$ mag, revealing the upper 2--3 magnitudes of the white
dwarf cooling sequence, and main sequence stars eight magnitudes and
more below the turn-off.  We fit the main sequence to subdwarfs of
known parallax to obtain a true distance modulus of $(m-M)_0 = 14.45
\pm 0.11$ mag.  A second distance estimate based on fitting the
cluster white dwarf sequence to field white dwarfs with known parallax
yielded $(m-M)_0 = 14.67 \pm 0.18$ mag.  We discuss the nature of the
difference between the two distance estimates and suggest approaches
for reducing the uncertainty in white dwarf fitting estimates for
future studies.  We couple our distance estimates with extensive
photometry of the cluster's RR Lyrae variables to provide a
calibration of the RR Lyrae absolute magnitude yielding $M_V(RR) =
0.42 \pm 0.10$ mag at [Fe/H] = --1.11 dex.  We provide another
luminosity calibration in the form of reddening-free Wasenheit
functions.  Comparison of our calibrations with predictions based on
recent models combining stellar evolution and pulsation theories shows
encouraging agreement, and the existing differences may provide useful
feedback to the models.

\end{abstract}


%%%%%%%%%%%%%%%%%%%%%%%%%%%%%%%%%%%%%%%%%%%%%%%%%%%%%%%%%%%%%%%%%%%%%%%%
\section{Introduction}  \label{sec_intro}

RR Lyrae variable stars (RRL) are among the most popular standard
candles for measuring distances to old stellar populations, both
within the Galaxy and to other galaxies in the Local Group.  Common
targets include Galactic globular clusters, the Galactic Center, the
Magellanic Clouds and the dwarf spheroidal companions of the Galaxy,
and even M31 and M33 and their companions.  The distance scale adopted
for the globular cluster system also defines the mean age and age
distribution of that system.  Globular clusters (GCs) place strong
constraints on the chronology of star formation and chemical evolution
in the early Galaxy.  Also, the age of the oldest GCs places a firm
lower limit on the age of the Universe, thus providing an important
consistency check on studies determining fundamental cosmological
parameters.  Clearly, an accurate calibration of the RR Lyrae absolute
magnitude, $M_V(RR)$, is vitally important for many fields of
astronomy.

\citet{cgcf00} reviewed the status of this calibration after
the results of the {\em Hipparcos} astrometry satellite had been
digested.  They compared the results of numerous distance estimation
techniques, and found encouraging evidence that the long-standing
dichotomy between the long and short distance scales was at last
yielding better consistency.  Still, evidence persisted that different
techniques produced systematically different results.  For example,
their careful analysis of main-sequence fitting techniques
consistently produced a distance scale about 0.1 mag brighter than the
mean of other methods.  They also demonstrated that the chief
limitation of this technique was the restricted sample of subdwarf
stars available with high quality parallaxes.  Until another
astrometry satellite is launched, it seems unlikely this technique
will undergo substantial improvement.

It is therefore worthwhile to explore and develop other techniques for
calibrating the RRL luminosity.  One technique still in its natal
stages is white dwarf (WD) sequence fitting.  WD fitting is analogous
to main sequence (MS) fitting except that the cluster's WD cooling
sequence is fit to local WDs with trigonometric parallaxes (e.g.,
\citet{renz96}), and/or theoretical model cooling sequences (e.g.,
\citet{wood95}).  WDs possess a number of advantages over MS stars
%which can result in more trustworthy fits.  First, WD atmosphere
%models in this temperature range are comparatively simple -- there is
%no convection involved, and the opacities are dominated by hydrogen or
%helium.  Thus the model cooling sequences are better defined than MS
%isochrones.  Second, the luminosity of a WD is nearly independent of
which may result in more trustworthy fits.  First, WD atmosphere
models in this temperature range involve different physics from MS
stars, some of which is simpler: there is no convection involved, and
the opacities are dominated by hydrogen or helium.  Complexities in WD
models such as core composition and crystalization should not produce
significant luminosity differences between WDs in globular clusters
and the field at the temperature and mass range of interest.  The
models thus involve a different set of systematic uncertainties with
respect to the MS models, and therefore provide a new and independent
means for obtaining globular cluster distances.  Second, the
luminosity of a WD is nearly independent of the star's initial, MS
composition.  Thus the large number of disk WDs near the Sun are
available as calibrators.  The main source of observational
uncertainty in WD fits involves WD masses, $m_{WD}$.  The
\citet{wood95} sequences indicate $\partial M_V / \partial m_{WD}
\approx 2.4$ mag~$m_\odot^{-1}$ at a given color \citep{renz96}.
Fortunately, GCs have a very narrow range of masses evolving off the
MS feeding the WD track, and various arguments indicate that $m_{WD} =
0.53 \pm 0.02~m_\odot$ for GCs \citep{rfp88,richer97}.  This
uncertainty translates into $\sim$0.05 mag in distance modulus.

Unfortunately, WDs are extremely faint, so WD fitting has not been
pursued extensively.  \citet{cool96} found the brightest WD in
NGC~6397, at $V \approx 23$ mag, $\sim$10 mag fainter than the
horizontal branch.  Accurate WD fits require reasonable photometry
($\sim$0.1 mag) several magnitudes down the WD sequence, so very
deep imaging is required. Significant WD sequences have been uncovered
in only a handful of globulars to date \citep{richer95, egs95, dmp95,
cool96, renz96, zoc01}.
%{richer95} Richer, H. B., et al. 1995, \apjl, 451, L17
%{egs95} Elson, R. A. W., Gilmore, G. F., Santiago, B. X., \&
%   Casertano, S. 1995, \aj, 110, 682
%{dmp95} Di Marchi, G. \& Paresce, F. 195, \aap, 304, 202
%{cool96}, {renz96}, {zoc01}
%(see \citet{cgcf00} for references).  
Most of these clusters have extremely red or blue horizontal branch
(HB) stars, and contain few or no RRL.  Any estimates of $M_V(RR)$
based on these clusters require extrapolating the HB across the RR
Lyrae pulsation strip.  This can be done with the aid of theoretical
HB models, but at the cost of the assumptions and uncertainties
inherent in those models.  The exception, M4, is rich in RRL, but
suffers from a high foreground reddening which varies spatially across
the cluster and which follows a non-standard reddening law
\citep{richer97}.  This complication translates into substantial
uncertainty in any RRL luminosity calibration thus derived.

The nearest globular with both a low reddening and abundant RRL is M5
(NGC~5904).  We have therefore undertaken a project to measure the
distance to M5 using WD fitting, and thereby calibrate the luminosity
of its RRL.  This paper describes our deep imaging of M5 using the
{\it Hubble Space Telescope} ({\it HST}), our
photometric analysis procedure and methodology for detecting WDs
(Sec. \ref{sec_obs}), and our distance analyses using both WD fitting
(Sec. \ref{sec_wdfit}) and main sequence fitting
(Sec. \ref{sec_msfit}).  In Secs. \ref{sec_rrlmag} and \ref{sec_mvrr},
we combine these results with extensive photometry of the M5 RRL
compiled from the literature to provide RRL luminosity calibrations in
several forms.  We discuss our results and how they might be improved
upon in Sec. \ref{sec_concl}.

%%%%%%%%%%%%%%%%%%%%%%%%%%%%%%%%%%%%%%%%%%%%%%%%%%%%%%%%%%%%%%%%%%%%%%% 
\section{Observations and Reductions} \label{sec_obs}
 
Images of a region in M5 were obtained using the {\it Wide Field and
Planetary Camera 2} (WFPC2) aboard {\it HST} as part of program
GO-8310.  The WFPC2 region is located at equatorial coordinates
$\alpha = 15^h 18^m 36.00^s$ and $\delta = +02^\circ 08^\prime
18.4^{\prime \prime}$ (J2000), about 200 arcsec North of the cluster
center at a point providing a high surface density of white dwarfs
with minimal crowding by bright cluster stars.

Images were obtained using the F555W and F814W filters during four
visits to the cluster as described in Table \ref{tab_visits}.  This table
includes the date of each visit, the orbit number within the visit,
the filter used, and the number and length of exposures obtained
during the orbit.  The telescope pointing was dithered slightly
between orbits two and three of Visits 1, 2, and 3, and between each
of the orbits of Visit 4.  Shorter exposures were included to provide
photometry of brighter stars useful for photometric comparisons with
ground based photometry.  The total exposure times in F555W and F814W
were 13,695 sec and 20,260 sec, respectively, obtained over fifteen
spacecraft orbits.

Raw images and optimal calibration frames were downloaded and
recalibrated ``on the fly'' from the Canadian Astronomical Data
Centre\footnote{{\tt http://cadcwww.dao.nrc.ca/}.}.  The resulting
calibration employed improved bias and dark files compared with the
original calibration provided by the Space Telescope Science
Institute.  
%The STSDAS task CRREJ was then used to create a cosmic ray rejection
%mask for each of the three long exposure images in a given orbit.

Rejection of cosmic rays was accomplished using the STSDAS task CRREJ.
In each orbit shown in Table \ref{tab_visits}, 
%the three long-exposure images were inspected for cosmic rays 
%by iteratively rejecting deviant
cosmic rays were detected in the stack of three long-exposure images
by iteratively rejecting deviant intensity values at each pixel
location, with the rejection criterion becoming stricter in each
iteration.  We also tested for cosmic rays in pixels adjacent to known
cosmic rays.  For each orbit, the result was a mask identifying the
location of cosmic rays on each of the three images.  The masked F555W
images from all six orbits were then shifted and combined using
MONTAGE2 \citep{turner96} to produce a high signal-to-noise image with
the cosmic rays rejected.  Note that in this image, the pixels where
cosmic rays were eliminated have a lower signal-to-noise than the
pixels which enjoyed uncorrupted data values.  In a typical mask for a
single image, we found that $\sim$6\% of the pixel locations were
affected by cosmic rays.  Using this cosmic ray rate, we performed a
simple statistical model to determine the fraction of WFPC2 pixel
locations affected, and to estimate the signal-to-noise degradation
caused by cosmic rays to our combined image.  About one third of the
WFPC2 pixels were unaffected by cosmic rays, while $\sim$38\% of the
pixel locations suffered a cosmic ray hit in one of the stacked
intensity values.  In total, $\sim$98\% of the WFPC2 pixels were
affected by three or fewer cosmic rays, causing a signal-to-noise to
drop of at most 9\% relative to pixel locations with no cosmic rays.
This indicates that the vast majority of the area surveyed has fairly
uniform characteristics that should not affect significantly our
detection of faint stars.  The worst case encountered in the model was
when seven cosmic rays hit a particular pixel location, leading to a
22\% drop in signal-to-noise relative to unaffected pixels.  However,
such cases were rare, affecting only $\sim$20 pixels across the entire
survey area.  We conclude that the presence of cosmic rays has had
minimal impact on the detection of faint stars.  As we will show in
the following section, detection incompleteness due to crowding is
significantly more important.

%%%%%%%%%%%%%%%%%%%%%%%%%%%%%%%%%%%%%%%%%%%%%%%%%%%%%%%%%%%%%%%%%%%%%%% 
%\section{The Color Magnitude Diagram}   \label{sec_cmd}
 
\subsection{Object Finding and Photometry}    \label{sec_cmd_phot}
 
We used SExtractor \citep{ba96} to find and classify sources on the
final, combined F555W image.  The resulting list of object positions
was used as input to ALLFRAME \citep{pbs94} to perform point spread
function (PSF) fitting photometry on the individual, cosmic ray-masked
F555W and F814W frames.  Each frame was first multiplied by a
geometric correction frame obtained from the {\it HST} Archive to
correct for distortions in the apparent area of pixels produced by the
WFPC2 cameras.  Experimentation with the tunable parameters in
ALLFRAME and SExtractor was done to optimize extraction of the faint
WD candidate stars in our frames.
%Both SExtractor and ALLFRAME were used
%in two passes -- first to find and photometer obvious objects, then in
%a second pass on the PSF-subtracted frames to find and photometer
%fainter, crowded objects that were missed in the initial object
%identification.  
The high signal-to-noise {\it HST} PSFs for F555W and F814W that were
utilized in the reduction of {\it HST} images by the Cepheid Distance
Scale Key Project team (Silbermann et al. 1996, kindly provided by
Peter Stetson) were applied to each of our 59 frames.  Positional
transformations between each frame and a reference image were used by
ALLFRAME to simultaneously iterate on the positions and magnitudes of
each object supplied by SExtractor.
%Given the positional transformation between each frame and a
%reference image, ALLFRAME was used to fit each PSF to a master object
%coordinate list derived using SExtractor. 
The result was profile-fitting photometry for approximately 8000
sources in each of the long exposure WF frames. We did not perform
photometry on the PC images since the small area of sky covered by that
chip would yield very few WD stars. 
%At this point, we followed the procedures outlined by \citet{raj00} to
%produce calibrated $VI$ photometry on the scale of \citet{silber96},
%which represents the work of the {\it HST} Cepheid Key Project.
 
Besides using SExtractor to find objects in the WFPC2 frames, we also
retained the ``class'' parameter computed by SExtractor.  While not
strictly a Bayesian classification, the value of class is
approximately the probability that any given object is stellar, with
1.0 representing definite unresolved sources, and 0.0 representing
definite non-stellar sources.  We will return to the use of this
morphological information below, when we discuss cleaning the
color-magnitude diagram.

We then reapplied SExtractor to find additional faint sources on the
PSF-subtracted images produced by ALLFRAME.  However, inspection of
the images showed that many detections were of
residual light where the PSF had been imperfectly subtracted from the
undersampled profile of an object detected in the first application of
SExtractor.  Furthermore, a high fraction (79\%) of these objects were
classified as non-stellar by SExtractor.  Though we ran these
additional objects through ALLFRAME to determine magnitudes for them,
further tests (see Sec. \ref{sec_cmd_clean}) indicated that they would
add to our data set many false detections and few stars upon which
high-quality photometry could be done.  Therefore, we report
magnitudes only for the objects detected in the first application of
SExtractor.  We note that the second application of ALLFRAME
photometry did help to improve the photometry of the stars found in
the first pass by accounting for, at least approximately, the light
from nearby, fainter objects.

\subsection{Calibrating the Photometry} 
 
%At this point, we followed the procedures outlined by \citet{raj00} to
%transform the instrumental photometry from ALLFRAME into calibrated
%$VI$ photometry on the scale of \citet{silber96}, which represents the
%work of the {\it HST} Cepheid Key Project. 

At this point, we followed the procedure outlined by \citet{raj00} to
obtain aperture corrections which convert our instrumental,
PSF-fitting photometry from ALLFRAME into the equivalent aperture
magnitudes via a zeropoint offset.  We then applied the
\citet{silber96} transformation equations to convert our instrumental
aperture magnitudes from the F555W and F814W filters into calibrated
$V$ and $I$ photometry (respectively) on the scale of
\citet{silber96}, which represents the work of the {\it HST} Cepheid
Key Project.

%Besides applying the \citet{silber96} transformations to the
%instrumental magnitudes, w
We then corrected for a number of well-known photometric effects in
the WFPC2 system.  As recommended by the WFPC2 Instrument Team,
time-dependent corrections for charge-transfer efficiency (CTE)
effects were installed based on the prescription of \citet{dol00},
updated using the data available on his web site\footnote{{\tt
http://www.noao.edu/staff/dolphin/wfpc2\_calib/}, updated 2002
Sept. 17.}.  The CCD dewar window throughput also changes as a
function of time.  After carefully examining the effect of adding
these small ($-$0.002 to +0.012 mag), photometric offsets, we
determined that the main sequences in the three different WF chips
were less aligned with the corrections than without, so we did not
apply these corrections.  Had we applied these corrections, their
effect would have been anyway small, with a maximum of +0.007 mag in
the $V$-band, and a maximum of +0.006 mag in $V-I$.

Table \ref{tab_phot} contains data on all the objects detected on the
three WF chips by the first application of SExtractor.
%Figure \ref{fig_cmd}a.  
The first column contains an identification number generated by
ALLFRAME, the second column indicates upon which chip the object fell
(WF2, WF3, or WF4), and the next two columns indicate the object's $X$
and $Y$ pixel coordinates on that chip.  Columns 5--8 show the $V$ and
$I$ magnitudes and their errors (specifically, the frame-to-frame
standard error of the mean, propagated through the photometric
transformation equations of \citet{silber96}).  The remaining columns
present the SExtractor parameters describing the ellipticity of the
object, its full-width at half-maximum in pixels, and the object's
morphological ``class'' value.  The electronic file contains data for
10,409 objects.

\subsection{Cleaning the CMD}   \label{sec_cmd_clean}
 
\begin{figure}
\plotone{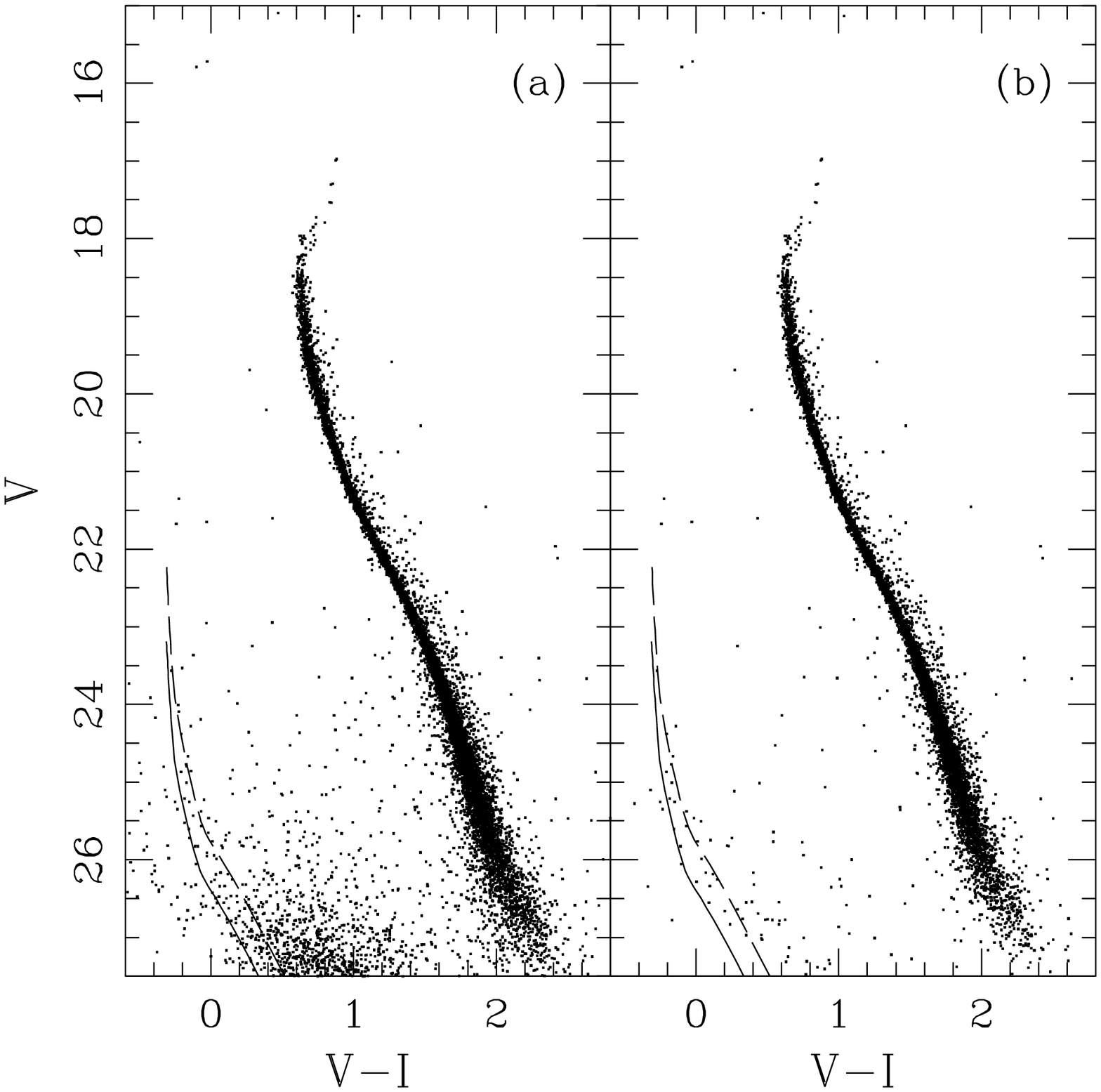}
\caption{The calibrated, $VI$ CMD for M5. (a) All objects from Table
\ref{tab_phot} are plotted as dots.  The dashed and solid lines
represent the WD cooling tracks of \citet{bwb95} for log(g) = 7.5 and
8.0 respectively, shifted to a distance modulus of $(m-M)_0 = 14.67$
mag and reddening of $E(V-I) = 0.046$ mag.  (b) Only objects
classified as stellar (see Sec. \ref{sec_cmd_clean}) are plotted.
\label{fig_cmd} }
\end{figure}

Figure \ref{fig_cmd}a shows the color-magnitude diagram (CMD) for all
objects listed in Table \ref{tab_phot}.  The curves represent the
log(g) = 7.5 and 8.0 WD cooling tracks from \citet{bwb95}, shifted to
account for typical reddening and distance modulus values for M5.
These tracks are meant only to guide the eye in finding the CMD
location of cluster WDs, and are not used in our WD fitting analysis
(Sec. \ref{sec_wdfit}).  Clearly, the large number of objects spread
throughout the middle and blue side of Figure \ref{fig_cmd}a makes it
difficult to determine whether WDs are present.  

%Some of these
%excess objects are image defects associated with the diffraction
%spikes and halos produced by bright stars, many are stars which are
%too crowded with neighboring stars to give precise photometry, while a
%few are background galaxies or field stars.

We employed the morphological ``class'' criterion of SExtractor to
statistically reject objects that did not have stellar profiles.
After experimenting with morphology cuts, we found that keeping only
objects with class $>$ 0.75 gave the greatest reduction in non-stellar
objects while retaining most stellar objects.  The exact cut value did
not matter greatly, and higher class values (greater probability of
the object being stellar) gave similar results.  The resulting CMD is
presented in Figure \ref{fig_cmd}b.

We note that the number of blue objects scattered between the WD and
MS regions in Figure \ref{fig_cmd}b would be much higher had we
included the objects detected in the second application of SExtractor
(to the PSF-subtracted image) discussed in Sec. \ref{sec_cmd_phot}.  A
CMD of these objects contains a much higher fraction of blue objects
relative to MS objects than is seen in Figure \ref{fig_cmd}b (29\% and
2\%, respectively), and only 1--2 objects fell near the \citep{bwb95}
WD tracks.  The vast majority of these sources were faint objects in
the wings of brighter stars that were found in the first application
of SExtractor.  Since our goal is to obtain the highest quality
photometry possible, and does not require a high degree of
completeness, our purposes are best served by omitting these stars
from the CMD.

The cleaned CMD shown in Figure \ref{fig_cmd}b has a narrow,
well-defined main sequence extending from just above the MS turn-off
to over eight magnitudes below the turn-off.  A few dozen WD
candidates can be seen in the region around the model cooling tracks.
This number agrees with expectations calculated from the number of
stars evolving off the MSTO in Figure \ref{fig_cmd}, by way of the
evolution rate determined from isochrones \citep{girardi02} and the WD
cooling rate \citep{bwb95}.  Such calculations are rough given the
small number of stars involved and the uncertain degree of detection
completeness suffered by the WD and MSTO stars.  Still, the agreement
between the observed and expected number of WDs in Figure
\ref{fig_cmd}b indicates that we are not missing a large fraction of
the WDs due to over-aggressive cleaning of the CMD.
%This number is in agreement with expectations based on the number
%of stars evolving off the MSTO in Figure \ref{fig_cmd}b in the context
%of evolution rates determined from isochrones \citep{girardi02} and WD
%cooling rates \citep{bwb95}.
%Girardi et al. 2000, A&AS, 141, 371
Most of the blue objects scattered between the WD and MS
regions in Figure \ref{fig_cmd}b can be attributed to distant,
unresolved galaxies.  Using statistics from the Hubble Deep Fields
\citep{hdf-n, hdf-s}, we expect there to be roughly 33 to 39
unresolved background galaxies with $V < 27$ mag in our CMD.  This
number is slightly less than the number of blue objects seen in Figure
\ref{fig_cmd}b, suggesting that the latter may include some 
field stars located behind the cluster in the Galactic halo.
%A few distant, unresolved galaxies may contribute to these 
%objects as well, especially at the faint end were the signal-to-noise
%could be too low for SExtractor to obtain a reliable ``class'' value.

% {hdf-n} Williams et al. 1996 AJ, 112, 1335
% {hdf-s} Casertano et al. 2000, AJ, 120, 2747

%Good WD candidates are available from
%just fainter than $V$ = 24 to the reliable limit of the photometry,
%perhaps $V \approx$ 27 mag, though we restrict our WD analysis to stars
%with $V < 26$ mag to minimize bias near the limit or our photometry.

\subsection{Final Photometry}    \label{sec_cmd_final}
 
The errors presented in columns 6 and 8 of Table \ref{tab_phot}
indicate the internal, random errors associated with our photometry.
We determined typical error values for stars in specific regions of
the CMD shown in Figure \ref{fig_cmd}b by computing the median
uncertainties in $V$ and $V-I$ within selected magnitude and color
ranges.  At $V \approx 21$ mag, the median errors in $V$ and $V-I$ are
0.015 and 0.019 mag, respectively.  They gradually increase at fainter
magnitudes, reaching 0.056 and 0.061 mag for stars at the lower end of
the MS, and 0.051 and 0.098 mag in the WD region.  The internal errors
for stars brighter than $V \approx 21$ mag increase to $\sim$0.036
and $\sim$0.047 mag because lower signal-to-noise photometry was
included from a few short exposure WFPC2 images.

To obtain an estimate of the external, transformation-based errors in
our photometry, we compared our final WFPC2 photometry with the M5
standard fields observed by \citet{pbs00}. For the fifteen bright
stars in common between the two datasets, we find offsets of $V_{us} -
V_{Stetson} = -0.053 \pm 0.010$ and $I_{us} - I_{Stetson} = -0.059 \pm
0.009$ mag, with no correlation with color or magnitude.  While the
zeropoint differences in the magnitude scales are statistically
significant, it is gratifying to find that the $V-I$ color scales are
essentially identical.  We note that these fifteen stars are among the
brightest in our data set.  They were derived from short exposure
images and may not accurately represent the photometry of WD and lower
MS stars, derived from the long-exposure images, about eight
magnitudes fainter.  Since it is not evident which zeropoints are
correct, we have chosen {\it not} to apply these magnitude offsets to
our photometry.  Instead, we will quote our distances with and without
these offsets.

It is instructive to obtain a sense of the depth of our photometry in
terms of the physical properties of low-luminosity stars.  We estimate
the mass of the faintest main sequence stars visible in our photometry
by comparing with the isochrones of \citet{girardi02}.  A slight
extrapolation below the low-mass limit of the isochrone with $Z =
0.001$ and age of 14.1 Gyr suggests the MS stars at our photometric
limit of $V = 27$ mag have masses of about 0.14 $m_\odot$.  Though our
photometry is deep, we are not reaching the MS hydrogen burning limit,
which is predicted by some models to be $\sim$8 mag below the faintest
MS stars in our data set.  Nor are we reaching the lower limit of the
WD cooling sequence expected at $V > 31$ mag, or the WD luminosity
function jump created by the changing atmospheric opacity due to
neutral hydrogen expected at $V \approx 29.5$ mag \citep{hansen04}.
While our data is not suited to examining the transition between lower
MS and brown dwarf stars or determining the age of M5 from its WD
stars, it is well suited for its intended purpose: a distance estimate
to M5 based on WD fitting.  
%Obtaining deeper photometry would require a
%much larger allocation of {\it HST} observing time.

%%%%%%%%%%%%%%%%%%%%%%%%%%%%%%%%%%%%%%%%%%%%%%%%%%%%%%%%%%%%%%%%%%%%%%% 
\section{White Dwarf Fitting Distance}    \label{sec_wdfit}
 
\begin{figure}
\plotone{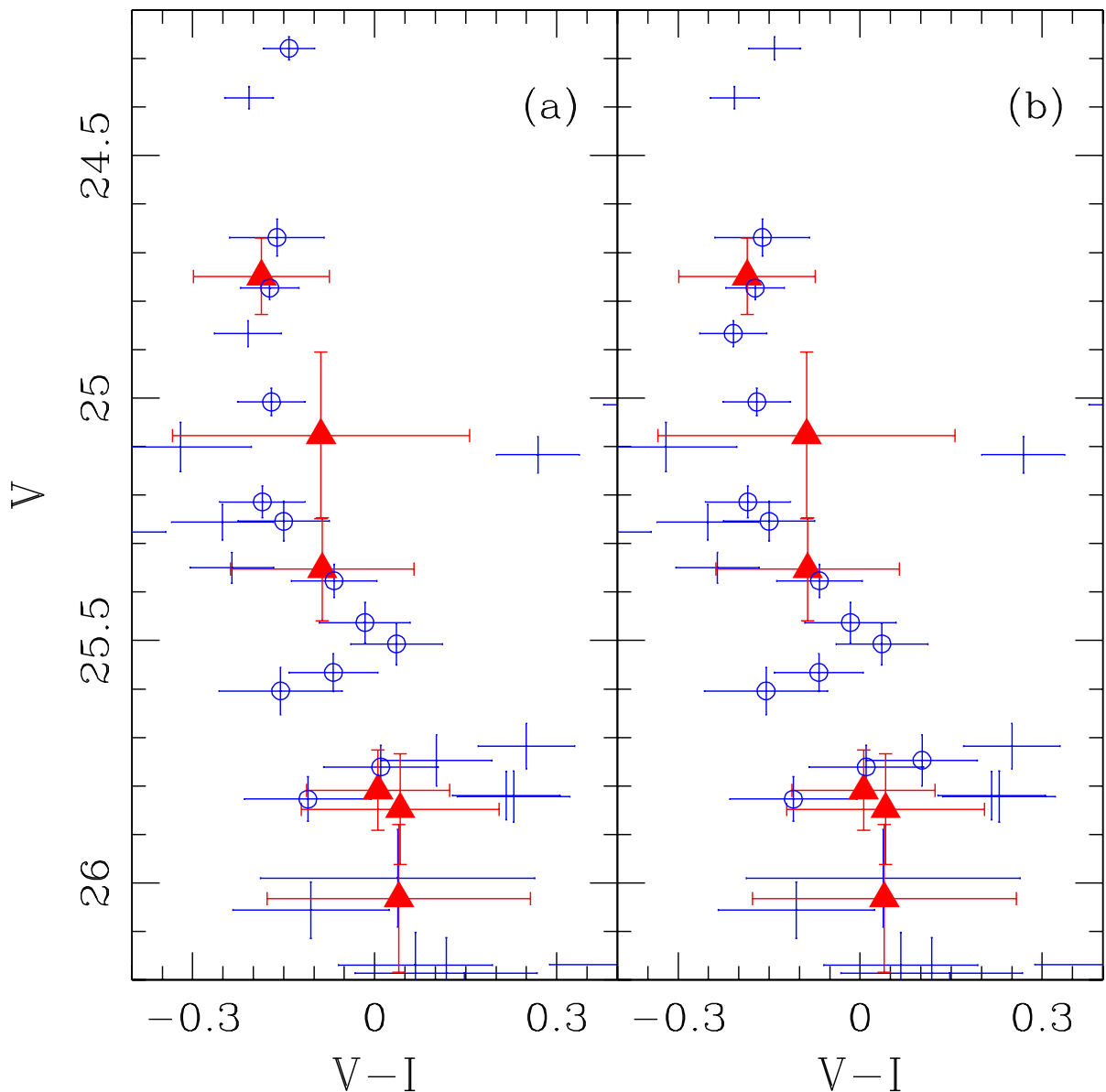}
%\plotone{f2.eps}
\caption{The WD region of the M5 CMD showing the WDs used (open
circles) and not used (bare error bars) in the distance fits in
Sec. \ref{sec_wdfit}.  The calibrating field WDs, shifted to the
fitted distance modulus of M5, are shown as triangles.  Panel (a)
shows the results of the first-pass fit, while (b) is for the
reselected WD candidates used in the second pass distance
determination. \label{fig_wdfit} }
\end{figure}

Figure \ref{fig_cmd}b presents a distinct sequence of several dozen WD
candidates running from $V \approx 24$ mag to the limit of the
photometry at $V \approx$ 27 mag.  We zoom in on the WD region in
Figure \ref{fig_wdfit}.  We restrict our WD distance analysis to stars
with $V < 26$ mag to minimize bias due to incompleteness near the
limit of our photometry, and to minimize uncertainty as the errorbars
increase with magnitude.  We inspected the F555W and F814W images at the
locations of the 27 cluster WD candidates in Figure \ref{fig_wdfit}.
Most candidates are isolated and should not be subject to systematic
photometry errors due to crowding.  There are several marginal cases
worth noting, however.  The star at ($V-I$, $V$) = (--0.32, 25.10) is
located near a brighter star whose diffraction spike probably
compromises our photometry.  The stars at (--0.21, 24.38), (+0.27,
25.12), and perhaps (--0.25. 25.26) may also have crowding-related
errors.  We placed less weight on these stars in our analysis as
described below.

\citet{zoc01} determined a distance to the globular cluster 47~Tuc by
matching WFPC2 photometry of local WDs which have known masses and
trigonometric parallaxes to WFPC2 photometry of likely WDs in 47~Tuc.
In order to avoid dependencies on the assumed photometric
transformations, Zoccali et al.  compared the local and 47~Tuc WD
samples in the instrumental magnitude system, after incorporating the
\citet{dol00} CTE corrections.  Zoccali et al. also corrected each of
the local WD calibrators to a WD mass of 0.53 M$_{\sun}$, the assumed
mass value of the WDs observed in 47~Tuc.  In order to be as
consistent as possible with their technique, we made a small
correction to their local WD instrumental magnitudes (kindly provided
by M. Zoccali) for the difference in CTE corrections they used and the
more updated CTE corrections currently supplied by Dolphin.  We
determined updated CTE corrections for the field WDs by downloading
the WFPC2 observations from the {\it HST} archives, then measuring the
WD fluxes and surrounding sky values for each frame.  The CTE
corrections under the two Dolphin prescriptions were computed and
compared to each other.  The difference in the two corrections only
amounts to 1\%, in the sense that the \citet{dol00} correction is 1\%
larger than the more modern correction, specifically 0.0013 mag
compared to 0.0003 mag, for these particular WDs.  Finally, we applied
the \citet{silber96} photometric transforms to the field WDs to put
them on the same scale as our M5 WDs.  This is equivalent to removing
the Silbermann et al.  photometric transform from the GC WDs and
determining distance in the instrumental magnitude system, as done by
Zoccali et al.  Table \ref{tab_wdphot} presents our final photometry
for the DA WDs found in Table 1 of \citet{zoc01}.
 
As done by \citet{zoc01}, a straight line was fit to the final
photometry for the field WD sample, and then this line was fit to the
GC WDs within the same color range ($V-I = -0.226$ to +0.002, i.e.,
without consideration of error bars).  This narrow requirement for
inclusion was chosen for this first iteration on determining the M5 WD
distance both to avoid extrapolation and to omit stars with
photometry possibly contaminated by crowding.
%  From that basic approach, our statistical technique diverged 
%somewhat from the technique employed by Zoccali et al.  
To accomplish the fitting, we used GaussFit, a code for least squares
and robust estimation \citep{jfm88}, which allowed us to fully
incorporate uncertainties in the colors and magnitudes, the covariance
between color and magnitude, and the uncertainty in the slope of the
fit to the field WD sequence.  The resulting distance modulus was
14.70 $\pm$ 0.12 mag.  In a slightly different approach, using the
same data but without the single most deviant (brightest) object, we
simultaneously fit a single slope and two intercepts for the field WD
and M5 WD data.  This yielded a distance modulus of 14.85 $\pm$ 0.32
mag.  Figure \ref{fig_wdfit}a presents the WD region of the CMD,
including the calibrating field WDs (triangles) and the M5 WDs used
(open circles) in these fits, and demonstrates why these two nearly
identical procedures might give significantly different uncertainties.
The error bars for the field WDs are large due to the combined
uncertainties in their photometry, trigonometric parallaxes, and the
corrections required to transform them to the comparison mass of 0.53
M$_{\sun}$.  The calibrating WDs also appear to form a steeper
sequence than the cluster WDs.  The steepness and scatter along the
sequence is most likely due to small differences in mass; either real
mass differences in the case of the M5 WDs, or small errors away from
the corrected mass values for the calibrating WDs.  It is also
possible that some of the M5 WDs are He-atmophere WDs, in which case
they should be 0.02 to 0.07 mag bluer \citep{bwb95} than the WD
calibrating sequence.
%Some He-atmosphere WDs may also be present among the M5 WD sequence.

We iteratively improved on this preliminary WD distance for M5 by
placing the calibrating WDs back in the CMD (see Figure
\ref{fig_wdfit}b) and reselecting the individual WD candidates.  This
time, instead of selecting cluster WDs based solely on their color, we
selected WDs based on both magnitude and color.  Specifically, we
included in the fits all WD candidates whose 1-$\sigma$ error ellipses
in the CMD overlapped with the sequence defined by the local WD
sample.  This two-parameter cut had the effect of dropping the
brightest WD candidate and including two other fainter WD candidates.
Figure \ref{fig_wdfit}b shows that it also selected against the stars
with questionable photometry.  The two statistical approaches tried
above (imposed slope and simultaneous fit) yielded distance moduli of
14.77 $\pm$ 0.11 and 14.79 $\pm$ 0.26 mag, respectively.  These
improved distances are within the errors of the distances derived
initially, and are slightly preferred on sample-selection grounds.

While the agreement of the two values is encouraging, the discrepancy
in the uncertainties highlights the sensitivity of the results to the
adopted fitting procedure.  As a conservative compromise, we adopt $(m
- M)_V = 14.78 \pm 0.18$ mag as our estimate of the distance modulus
for M5 as derived from its WD stars.

%%%%%%%%%%%%%%%%%%%%%%%%%%%%%%%%%%%%%%%%%%%%%%%%%%%%%%%%%%%%%%%%%%%%%%% 
\section{Main-Sequence-Fitting Distance}     \label{sec_msfit}
 
Our deep photometry of M5 provides us with an opportunity to determine
the cluster's distance by fitting its MS to nearby subdwarfs with
known trigonometric parallaxes.  A number of inputs are required,
including a fiducial sequence for the MS region of the cluster, a set
of subdwarf stars with known metallicities and trigonometric
parallaxes, and a prescription for adjusting the colors of the
subdwarfs for the effects of metallicity.
 
\begin{figure}
\plotone{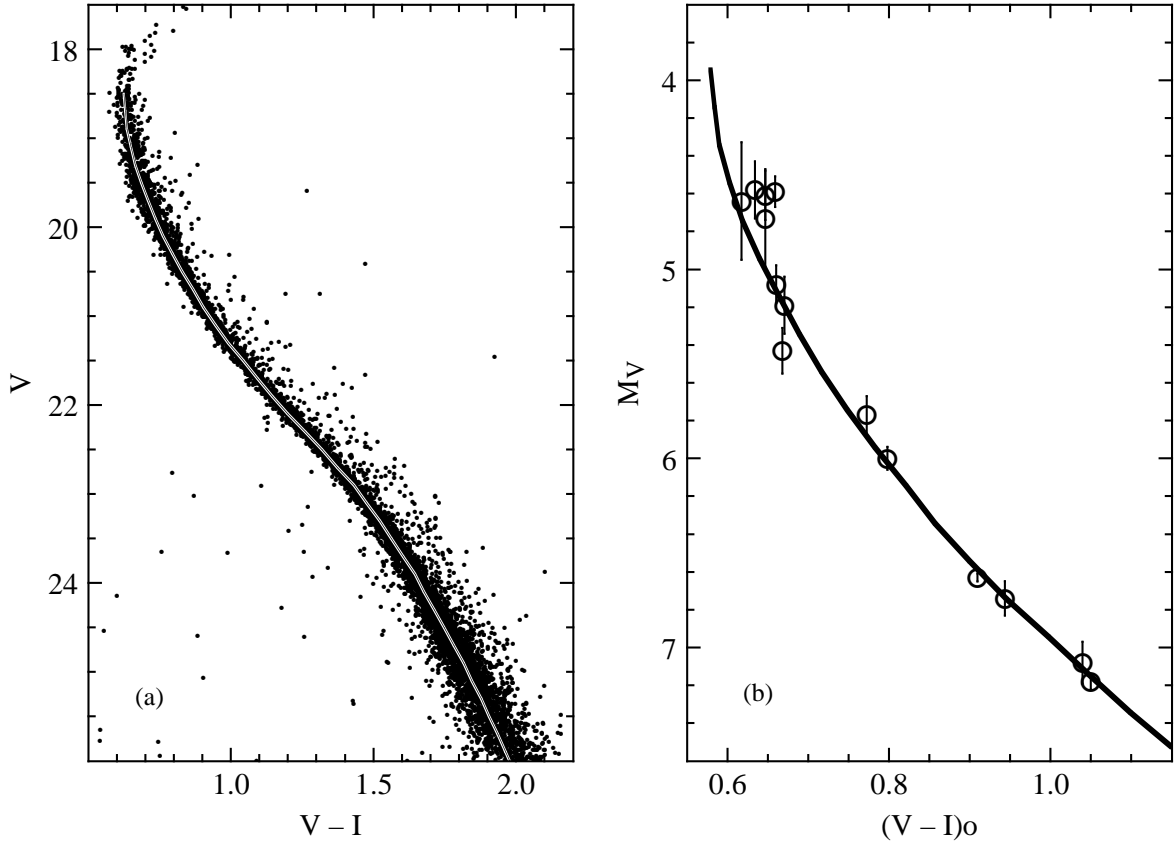}
\vskip -2.0in
%\caption{The weighted least squares fit of the M5 fiducial main
%sequence (solid line) to the field subdwarfs with {\it Hipparcos}
%parallaxes (open circles).  The adopted reddening and metallicity are
%indicated, along with the derived distance modulus.  \label{fig_sdfit}
%}
\caption{The CMD in panel (a) shows the M5 fiducial main sequence
plotted over the stars.  In panel (b), we show the weighted least
squares fit of the M5 fiducial main sequence (solid line) to the field
subdwarfs with {\it Hipparcos} parallaxes (open circles).  The adopted
reddening and metallicity, along with the derived distance modulus,
are discussed in Sec. \ref{sec_msfit}.  \label{fig_sdfit} }
\end{figure}

The main sequence fiducial has been constructed in the following
manner.  We divided the data into bins of 0.2 mag between $V = 18.5$
and $V = 26.1$ mag.  Within each bin, we compute the median color of
all stars and then perform a 1$\sigma$ rejection until the median
color difference from iteration to iteration is less than 0.005
mag. The resultant fiducial sequence, shown in Figure
\ref{fig_sdfit}a, does not appear to be influenced by the unresolved
binaries which lie above and to the right of the MS.
 
The set of subdwarf stars has been taken from the work of
\citet{sandq99} as listed in their Table 4. All of these stars possess
{\it Hipparcos} parallaxes and their absolute magnitudes have been
corrected for the Lutz-Kelker bias as described in
\citet{sandq99}. All but one of these stars has a more recent
metallicity determination from the work of \citet{gcclb03}. The mean
difference in metallicity is $0.09 \pm 0.03$ dex in the sense
(Sandquist -- Gratton). For the one star without a Gratton et
al. metal abundance measurement (BD~+54~1216), we offset the Sandquist
et al. value by --0.09 dex to account for the different abundance
scales.
 
The colors of the subdwarf stars are adjusted for their metallicity
using the \citet{girardi02} isochrones for Z=0.00001, 0.0004,
0.001, 0.004, 0.008, and 0.019 in the range $4.5 < M_V < 8.0$ mag. We
have parameterized the $V-I$ colors of these models along the MS as a
function of $M_V$ and [Fe/H]. Then, the partial derivative of this
relation ($\partial (V-I) / \partial$[Fe/H]) is used to adjust the
subdwarf colors to the metallicity of M5, which we take to be [Fe/H]$
= -1.11 \pm 0.03$ \citep{cg97}.
 
With the above-mentioned points in mind, and adopting a reddening to
M5 of $E(V-I) = 0.046 \pm 0.020$ mag (see Sec. \ref{sec_mvrr}), we
performed a weighted least squares fit of the M5 MS fiducial to the
metallicity-adjusted subdwarf photometry taken from Table 4 of
\citet{sandq99}. The resultant fit, shown in Figure \ref{fig_sdfit}b,
yields an apparent $V$-band distance modulus of $(m-M)_V = 14.56 \pm
0.10$ mag.  The error was computed from the uncertainty due to
reddening ($\sim$0.10 mag), the standard error of the subdwarfs around
the fitted fiducial ($\sim$0.01 mag), and the effect of a random
metallicity error of 0.03 dex on the results of the fit ($\sim$0.01
mag).  Correction for interstellar extinction results in a true
distance modulus of $(m-M)_0 = 14.45 \pm 0.11$ mag, which includes the
additional error inherent in $A_V$.

Our value for the distance modulus of M5 is in good agreement with the
MS fitting result of \citet{testa04}, who found $(m-M)_0 = 14.44 \pm
0.09$ (random) $\pm 0.07$ (systematic) mag.  Other recent main
sequence fitting results for M5 yielded true distance moduli of $14.46
\pm 0.05$ mag \citep{cgcf00}, $14.52 \pm 0.15$ mag \citep{reid98}, and
$14.42 \pm 0.09$ mag \citep{cdkk98}.  Much of the uniformity between
these results arises from the sample of subdwarfs employed, which is
defined by the availability of parallaxes from the {\it Hipparcos}
satellite.  Subtle differences between them include the photometric
zeropoints, which subdwarfs were used, and the way in which the colors
of the subdwarfs were corrected for metallicity effects.  If we apply
the offset to our photometry indicated by \citet{pbs00} ground-based
photometry (see Sec. \ref{sec_cmd_final}), our value becomes $(m-M)_0 =
14.50 \pm 0.11$ mag.

%%%%%%%%%%%%%%%%%%%%%%%%%%%%%%%%%%%%%%%%%%%%%%%%%%%%%%%%%%%%%%%%%%%%%%% 
\section{Apparent Magnitudes of the RR Lyrae}     \label{sec_rrlmag}

Having established estimates for the distance modulus to M5, we can
provide a calibration of the RR Lyrae absolute magnitude, $M_V(RR)$.
This requires a careful measurement of the mean apparent $V$ magnitude
of the ensemble of RRL in M5, $V(RR)$.  The RRL themselves do not
appear in our {\em HST} data, so we rely on ground-based observations
compiled from the literature.  Photometry in the $B$, $I$, and $K$
passbands are also available in the literature, enabling us to
calibrate the RRL absolute magnitude in these bandpasses as well.  We
note that systematic zeropoint differences may exist between these
sources and our photometry (in Table \ref{tab_phot}, or if corrected to
the \citet{pbs00} system).  Such problems are common in $M_V(RR)$
calibrations like MS fitting where the deep MS photometry and shallow
time-series photometry of the RRL are often obtained by different
researchers using different detectors, standard stars, etc.

\begin{figure}
\plotone{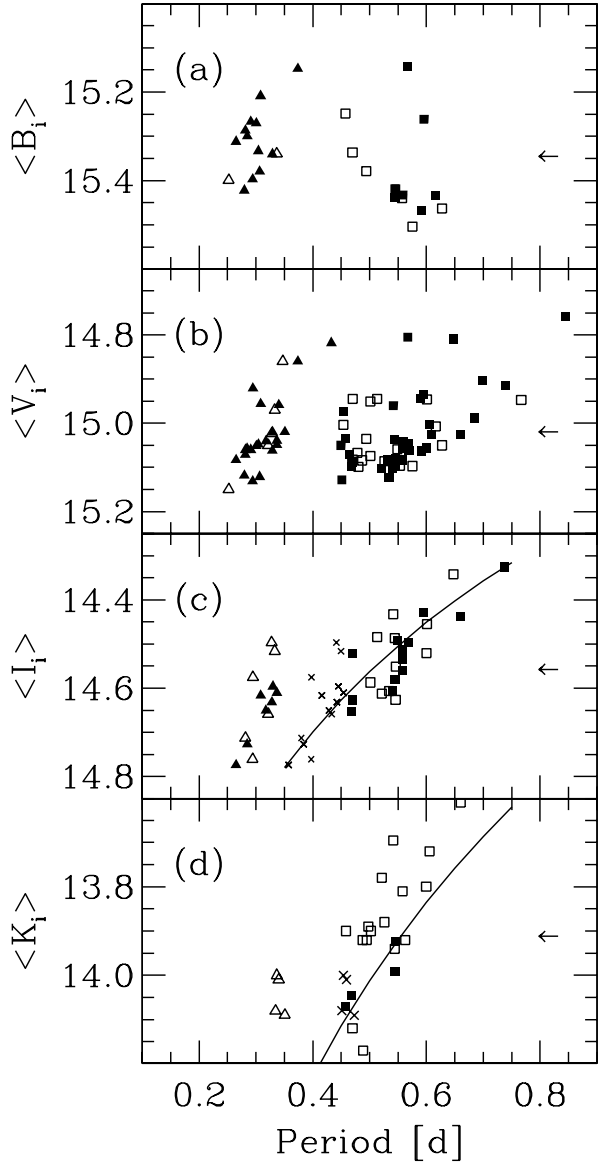}
\caption{Apparent magnitudes of RRL in M5 compiled from the
literature, plotted as a function of pulsation period, for the $B$,
$V$, $I$, and $K$ passbands.  In each figure, the arrow indicates the
mean apparent magnitude of the RRL. Squares and triangles indicate
RRab and RRc pulsators, respectively.  Solid symbols indicate stars
with light curves having complete phase coverage, while open symbols
indicate stars with small phase gaps.  The crosses in (c) and (d)
indicate the fundamentalized positions of the RRc stars, while the
curves are the best fits represented by the coefficients in lines 1
and 2 of Table \ref{tab_wasen}.  \label{fig_rrlmag} }
\end{figure}

Light curves for RRL in M5 are presented in several studies, including
Brocato, Castellani \& Ripepi (1996, $B$ and $V$ magnitudes); Caputo
et al. (1999, $B$ and $V$); Cohen \& Matthews (1992, $K$); Kaluzny et
al. (1999, $V$); Longmore at al. (1990, $K$); Reid (1996, $V$ and
$I$); Storm, Carney \& Beck (1991, $B$ and $V$); and Storm, Carney \&
Latham (1992, $K$).  To ensure self-consistency of our statistics, we
took the light curves of each star from their original source and
recomputed the star's intensity-mean magnitude, $\langle m_i \rangle$,
separately in each filter ($B$, $V$, $I$, and $K$).  This entailed
converting the individual magnitude estimates in a light curve into
intensity estimates, integrating under the phased light curve, and
converting the resulting mean intensity back into a magnitude.  We
rejected any star with a phase gap large enough to bias the estimate
of $\langle m_i \rangle$.  The resulting intensity-mean values are
shown in Figure \ref{fig_rrlmag} as a function of each star's
pulsation period.

Four studies provide $V$ magnitudes.  We compared the $\langle V_i
\rangle$ estimates for stars common to different pairs of studies to
search for any systematic differences in photometric zero-points
between the studies.  Differences of order 0.02 mag were found, but
they appeared to reflect the sub-samples of stars involved more than
systematic photometric differences.  We therefore took the $\langle
V_i \rangle$ values at face value.  If, for a given star, $\langle V_i
\rangle$ values were available from more than one study, we averaged
the values together to get a mean value for the star.  In this way, we
compiled a list of 79 RRL including 53 RRab and 26 RRc.

Figure \ref{fig_rrlmag}b shows the $\langle V_i \rangle$ value for
each star plotted against the star's period.  Statistics on this
sample is shown in the first row of Table \ref{tab_rrlmag}.  The
columns in this table give the number of stars used in the sample
($N_{RR}$), the arithmetic mean of the $\langle m_i \rangle$ values
$m(RR)$, the standard error of the mean ($sem$), the standard
deviation about the mean ($\sigma$), and the median of the $\langle
m_i \rangle$ values ($median$).  The table presents separate
statistics for RRab and RRc variables, though in $V$ there is no
significant difference between the mean magnitudes of the RRab and RRc
stars.  We adopt $V(RR)= 15.024 \pm 0.009$ mag as the mean apparent
magnitude of the 79 RRL studied in M5.  The error estimate reflects
the observed scatter in the stellar magnitudes, while a systematic
uncertainty of $\lesssim$0.02 mag can be expected in the photometric
zeropoint.  For comparison, \citet{harris96}\footnote{Data taken from
the 1999 June 22 update available at {\tt
http://physun.physics.mcmaster.ca/\~harris/mwgc.dat}.} lists the HB
magnitude of M5 to be 15.07 mag.  The lower envelope of the
distribution of the points in Figure \ref{fig_rrlmag}, which should
correspond to the Zero-Age Horizontal Branch locus, is at $15.10 \pm
0.04$ mag.

We compiled $B$-band data from the three sources listed above.  There
were no stars in common between the three studies, so no star-by-star
analysis was possible and no systematic zero-point corrections were
made.  Figure \ref{fig_rrlmag}a shows the $\langle B_i \rangle$ value
for each star plotted against the star's period.  The statistics for
RRab and RRc considered separately (see Table \ref{tab_rrlmag})
indicate that the RRc stars in M5 are significantly brighter than the
RRab stars.  The mean apparent $B$-band magnitude of the combined
sample of RRL in M5 is $B(RR) = 15.350 \pm 0.018$ mag.  As for $V$,
the error estimate reflects the observed scatter, while a systematic
error of $\lesssim$0.02 mag is possible in the photometric zeropoint.

Light curves in the $I$-band are available only from one source
\citep{reid96}.  The statistics shown in Table \ref{tab_rrlmag}
indicate that the RRc stars are significantly fainter in $I$ than the
RRab stars, i.e., in the sense opposite to that of the $B$ filter.
Furthermore, Figure \ref{fig_rrlmag}c shows a strong correlation
between $I$ magnitude and period among the RRab stars.  This
period-luminosity correlation is well-known in redder passbands such
as $K$ (e.g., Longmore et al. 1990).  To characterize this
correlation, we fundamentalized the periods of the RRc stars
(log~$P_f$ = log~$P_{RRc} $ + 0.13, \citet{cq87}), and performed a
least squares fit to all the RRL data using an equation of the form
\[I(RR) = a + b~({\rm log}~P_f + 0.3).\]  
Line 1 of Table \ref{tab_wasen} presents the fitted coefficients and
their uncertainties, the rms scatter of the points about the fit, and
the number of points used in the fit.  This relation may be subject to
a systematic zeropoint uncertainty of $\lesssim$0.02 mag.

Light curves in the $K$-band are available from three studies.  Both
\citet{cm92} and \citet{storm92} provided well-sampled light curves
for small numbers of stars (four and two, respectively) calibrated to
the CIT standard system of \citet{elias82}.  \citet{longm90} took the
opposite approach, obtaining only 1--2 observations per star for 23
separate stars.  This yields a larger sample of stars, but with
increased scatter (rms $\approx$ 0.05 mag) around a true mean
relationship due to phase-sampling effects.  Also, the \citet{longm90}
data are not directly calibrated to the \citet{elias82} standards,
though the authors argue there should be no systematic offset.  Figure
\ref{fig_rrlmag}d shows the $\langle K_i \rangle$ value for each star
plotted against the star's period, and Table \ref{tab_rrlmag} provides
statistics.  There is evidence for discrepancies in the photometric
zeropoints of the three studies at the 0.05--0.10 mag level, though
there are too few stars in common to determine any such shifts with
accuracy.  Normally, the $K$-band luminosity of RRL is characterized
as a function of period.  The scatter in the \citet{longm90} data
results in uncertainty in the period-dependence derived from Figure
\ref{fig_rrlmag}d.  To reduce this uncertainty, we adopt the mean
period-dependence (gradient) of the eight clusters presented in Table
4 of \citet{longm90}, and fit it to the $\langle K_i \rangle$ data for
the four stars with complete light curves using an equation analogous
to the one for $I(RR)$ above.  Line 2 of Table \ref{tab_wasen} shows
the results of the fit.  We estimate the systematic uncertainty in the
zeropoint of this relation to be $\sim$0.07 mag.  Clearly,
well-sampled $K$-band light curves for many RRab and RRc in M5 are
needed to ensure a definitive $K$-band luminosity calibration for M5.

%%%%%%%%%%%%%%%%%%%%%%%%%%%%%%%%%%%%%%%%%%%%%%%%%%%%%%%%%%%%%%%%%%%
\section{RR Lyrae Luminosity Calibration} \label{sec_mvrr}
 
To obtain a final calibration of the RR Lyrae absolute magnitude,
$M_V(RR)$, we must combine our distance modulus and apparent
magnitudes with interstellar extinction information.  From the WD
distance analysis, we obtained an apparent distance modulus of
$(m-M)_V = 14.78 \pm 0.18$ mag, while the main sequence fitting
analysis yielded $(m-M)_V = 14.56 \pm 0.10$ mag.  We are encouraged by
the agreement of the results relative to their formal errors.  We
combined the two results via a weighted mean, using the
inverse-squared errors for the weights, to obtain a single estimate of
the apparent distance modulus toward M5: $(m-M)_V = 14.61 \pm 0.09$
mag.  Clearly, the main sequence fit dominates the combined result.

\citet{testa04} advocated the reddening value $E(B-V) = 0.035 \pm
0.005$ mag for M5.  We adopt this value along with a more conservative
uncertainty of 0.01 mag, which we believe provides a more realistic
assessment of the systematic zeropoint uncertainty in the reddening
scale (see Sec. 7.4 of Schlegel et al. 1998).  This leads to a true
distance modulus of $(m-M)_0 = 14.50 \pm 0.10$ mag.  We note that the
main sequence fitting result is subject to a small systematic
uncertainty due to the zeropoint disagreement between our photometry
and that of \citet{pbs00}.  The WD value was obtained using photometry
of the field and M5 WDs on the {\it HST} instrumental system, so
should not suffer from this uncertainty.  If we shift our photometry
to match Stetson's calibration, the combined distance modulus becomes
$(m-M)_0 = 14.54 \pm 0.10$ mag.

Using the relations of Cardelli et al. (1989, their Table 3) in
conjunction with our adopted value of $E(B-V) = 0.035 \pm 0.010$ mag,
we determine the interstellar extinction values to be $A_B = 0.145 \pm
0.044$ mag, $A_V = 0.109 \pm 0.031$ mag, and $A_K = 0.012 \pm 0.004$
mag.  We note that the $I$ filter referred to in \citet{ccm89} has a
longer effective wavelength than the Kron-Cousins $I$ filter, to which
the {\it HST} photometric transformations are calibrated
\citep{silber96}.  Using transmission curves of the CTIO $I_{KC}$
filters,\footnote{See {\tt
http://www.ctio.noao.edu/instruments/filters/index.html}.} we obtain
$A_I = 0.58 A_V$ and therefore $A_I = 0.063 \pm 0.018$ mag in the case
of M5.  Note that this relation yields $E(V-I) = 1.30 E(B-V)$, in good
agreement with the Kron-Cousins reddening relation described by
\citet{dwc78}.  We adopt an increased uncertainty in $E(V-I)$ of 0.02
mag to reflect the uncertainty in the reddening relations.

Combining these extinction values with the apparent magnitudes from
Table \ref{tab_rrlmag} and the true distance modulus, we obtain the
absolute magnitudes given in the last column of Table
\ref{tab_rrlmag}, $M(RR)$.  We include separate values for RRab and
RRc stars.  The calibrations corresponding to lines 1 and 2 of Table
\ref{tab_wasen}, 
\[M(RR) = a + b~({\rm log}~P_f + 0.3),\]
are shown in lines 3 and 4 of that table, in which the quoted errors
include the systematic uncertainties due to the distance modulus and
to the photometric zeropoints discussed in Sec. \ref{sec_rrlmag}.  The
$M(RR)$ values in Table \ref{tab_rrlmag} become 0.04 mag brighter if
we adopt the \citet{pbs00} photometric system.

Another approach to representing the luminosity of RR Lyrae stars is
through the Wasenheit functions (e.g., Kovacs \& Walker 2001, Cassisi
et al. 2004, and references therein).  This reddening-free quantity is
especially useful in regions where the reddening is high and varies on
small spatial scales.  Our high quality distance modulus and RR Lyrae
photometry for M5 provide an opportunity to derive an empirical
calibration of the Wasenheit functions.  This is of particular
interest since theoretical calibrations have recently become available
through the evolution and pulsation modeling of \citet{cassisi04}.
Additional empirical calibrations may provide useful feedback to the
models.

We follow the studies mentioned above in defining the Wasenheit
functions $W(BV) = V - R_V(B-V)$ and $W(VI) = V - R_I(V-I)$, where $R$
is the ratio of total absorption to color excess in the appropriate
passbands.  While both studies adopt $R_V = 3.10$, \citet{kw01} employ
$R_I = 2.5$, while \citet{cassisi04} use 2.54, and we obtain 2.39 from
our analysis of the CTIO $I_{KC}$ filter transmission curves.  We also
define $W(VK) = K - R_K(V-K)$ with $R_K = 0.13$ for use with our
compiled $K$-band data set.  Each of these apparent values can be
converted to an absolute one via our true distance modulus, $W_0 = W -
(m-M)_0$.

\begin{figure}
\plotone{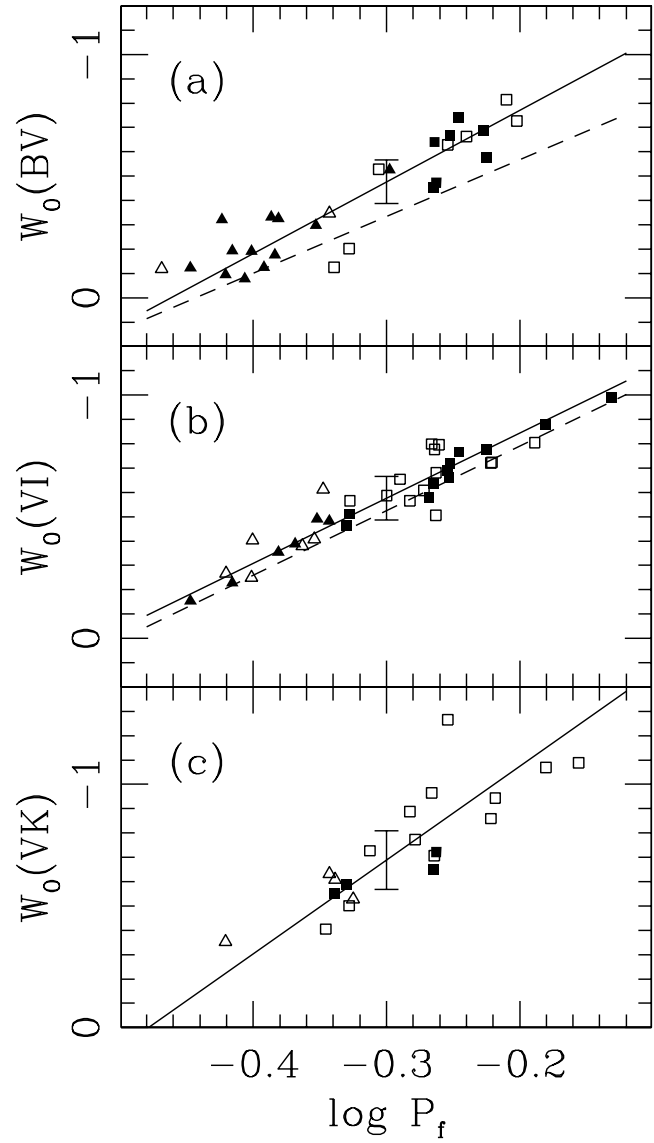}
\caption{Wasenheit functions for the M5 RRL as a function of
fundamentalized period.  The symbols are as in Figure
\ref{fig_rrlmag}.  Solid lines represent our least-squares fits to the
data, while dashed lines indicate the predictions of models by
\citet{cassisi04}.  Error bars indicate systematic uncertainty in the
fit due to uncertainties in the distance modulus and photometric
zeropoints.  \label{fig_wasen} }
\end{figure}

Figure \ref{fig_wasen} shows the $W_0$ values for our compiled RRL
data plotted as a function of fundamentalized pulsation period.  The
solid line in each panel indicates the least-squares fits to the data
in the form
\[ W_0 = a + b~({\rm log}~P_f + 0.3). \]
The fitted coefficients and their errors are given in Table
\ref{tab_wasen} along with the rms scatter of the points about the
fit, the number of points in the fit, and the adopted value of $R$.
The error in the coefficient $a$ is a quadratic combination of the
error in the fit (typically 0.01--0.03 mag) and the error in the M5
distance modulus (0.10 mag).  The values of the $a$ coefficient would
become 0.04 mag smaller if we employed the distance modulus
appropriate to the \citet{pbs00} photometric zeropoint.  Notice that
in each panel of Figure \ref{fig_wasen}, the stars with complete light
curves (solid points) cluster around the fitted line, while stars with
phase gaps in their light curves (open symbols) are more often
outliers.

In panels (a) and (b), the dashed lines indicate the relation
predicted by \citet{cassisi04} for their model having metallicity ($Z
= 0.001$) and horizontal branch type (+0.11) closest to the observed
values for M5.  For $W_0(BV)$, our slope is significantly steeper than
the predictions (2.5$\sigma$), while our zeropoint is only 1.6$\sigma$
brighter than the predictions owing to systematic uncertainty in the
level of the points due to the distance modulus uncertainty.  The
star-to-star dispersion is also larger than predicted, suggesting that
there may be more star-to-star variation in properties such as stellar
mass than is expressed in the models.  We note that the observational
work of \citet{kw01} found a slope of --2.47, in better agreement with
the models.

For $W_0(VI)$, the agreement between our observations and predictions
appears better (see Figure \ref{fig_wasen}b), with the slope and
zeropoint deviating by 0.2$\sigma$ and 0.6$\sigma$, respectively.  If
we adopt $R_I = 2.54$ to match \citet{cassisi04}, the points in Figure
\ref{fig_wasen}b shift upward and the slope steepens, giving the
fitted coefficients shown in line seven of Table \ref{tab_wasen}.
Still, the agreement is better than with $W_0(BV)$, showing
differences from the predictions by only 1.1$\sigma$ and 1.3$\sigma$
for the slope and zeropoint, respectively.  The scatter is also
smaller, in better agreement with the predictions.  The slope of the 
observed RRL becomes steeper if we convert our intensity magnitudes to
the static magnitudes used by the models \citep{marconi03}.
%{marconi03}Marconi etal (2003, ApJ, 596, 299)

\citet{cassisi04} discuss the strengths and weaknesses of their
models.  They note that the slope predictions derive from the relation
between pulsation period and stellar luminosity and effective
temperature.  These relatively well-established elements of the
modeling should result in slope predictions that are reliable.
Meanwhile, the predicted zeropoints depend on elements of the models
that are less constrained (the luminosity of the horizontal branch
models and the bolometric corrections), so are less secure.  It
appears that the input parameters to the models, or perhaps the models
themselves, could be adjusted to provide steeper slopes and brighter
zeropoints to provide a better match to our observations of M5.

%%%%%%%%%%%%%%%%%%%%%%%%%%%%%%%%%%%%%%%%%%%%%%%%%%%%%%%%%%%%%%%%%%%
\section{Conclusion}      \label{sec_concl}
 
We have presented deep {\it HST} observations of the globular cluster
M5.  The data reach over eight magnitudes below the main sequence
turn-off and include a number of cluster white dwarf stars.  By
fitting our deep main sequence to a sample of subdwarfs having
trigonometric parallaxes, we obtain an apparent distance modulus of
$(m-M)_V = 14.56 \pm 0.10$ mag, in good agreement with other main
sequence fitting solutions for M5.  If the true metallicity of M5 is
different from the adopted value of [Fe/H] = --1.11 dex, the distance
modulus shifts systematically by 0.4 mag/dex.

We also described our approach for selecting a sample of WDs with the
best quality photometry, and for obtaining a distance estimate to the
cluster based on comparison with field WDs having trigonometric
parallaxes.  The resulting distance modulus, $(m-M)_V = 14.78 \pm
0.18$ mag, is in good agreement with other estimates for M5,
indicating that our method is reliable.  However, the uncertainty in
our WD distance modulus is large compared with the uncertainties
associated with more established methods like main sequence fitting.

Much of the uncertainty in our WD distance modulus is attributable to
the extreme depth of the observations ($V \approx 25$ mag) in the WD
region, and the correspondingly low signal-to-noise of our WD
magnitude and color measurements.  Newer instruments on {\it HST}
provide better throughput than WFPC2, so could provide higher
signal-to-noise at this magnitude.  We also note that $V-I$ does not
offer the best WD sequence for distance estimation; bluer filters and
longer color baselines would result in a WD sequence with a shallower
slope.  However, the relatively low blue throughput of WFPC2 demanded
that we work in the F814W (``$I$'') filter for our study.  Another
advantage the newer {\it HST} instruments have over WFPC2 is their
superior spatial resolution, which would provide SExtractor with more
information to separate WDs from critically resolved, blue background
galaxies.  Three filter photometry would also help in this separation,
as it would with separation of cluster WDs from blue field subdwarf
stars.  Finally, a second visit to M5 using WFPC2 or one of the
advanced imagers on {\it HST} would provide both higher
signal-to-noise magnitudes, and an opportunity to reject field stars
and background galaxies using proper motion information
\citep{king98}.  Observations of additional fields in M5 would
provide a larger sample of WDs, which would also lead to an improved
WD-based distance estimate.

We combined our main sequence and WD results to obtain a best estimate
for the true distance modulus of M5: $(m-M)_0 = 14.50 \pm 0.09$ mag.
This result is weighted strongly toward the main sequence fit result.
Using this in conjunction with a large sample of RR Lyrae magnitudes
compiled from the literature, we obtained a calibration for the RR
Lyrae absolute magnitude $M(RR)$ in the $B$, $V$, $I$, and $K$
passbands (see Tables \ref{tab_rrlmag} and \ref{tab_wasen}).  Our
value of $M_V(RR) = 0.42 \pm 0.10$ mag is brighter than the value of
$0.64 \pm 0.07$ mag advocated by \citet{cgcf00} for [Fe/H] = --1.11
dex.  This is consistent with their finding that main sequence fitting
calibrations tend to be about 0.1 mag brighter than the average value
derived from many independent techniques.  The distance modulus
obtained from our WD fit alone would yield $M_V(RR)$ farther from the
\citet{cgcf00} value.

We also presented period-luminosity calibrations in the form of
reddening-free Wasenheit functions.  These are useful both for
distance estimation to heavily reddened systems and for comparison
with predictions of recent models that combine evolution and pulsation
theories \citep{cassisi04}.  Our empirical calibration suggests that
the current models predict a period dependence that may be too weak
and a luminosity zeropoint that may be too faint.  However, the
systematic over-brightness of main sequence fitting \citep{cgcf00}
tends to compensate, bringing the theoretical and observed zeropoints
into closer agreement.

We conclude that large aperture telescopes with blue-sensitive imagers
will enable the WD fitting technique to become increasingly valuable
in helping to refine the RR Lyrae luminosity calibration, both for M5
and other globular clusters rich in RR Lyrae stars.

%%%%%%%%%%%%%%%%%%%%%%%%%%%%%%%%%%%%%%%%%%%%%%%%%%%%%%%%%%%%%%%%%%%%%%%%
\acknowledgments

The authors thank an anonymous referee and Marco Castellani for
valuable comments and suggestions.  We thank Peter Stetson for kindly
providing us with PSFs for {\it HST} F555W and F814W images, Manuela
Zoccali for providing us with her instrumental photometry of local
field WD stars, Barbara McArthur for help with GaussFit, and Andrew
Dolphin for providing CTE correction updates on his website.  We also
thank Janusz Kaluzny, Vincenzo Ripepi, and Jesper Storm for sending
electronic files containing the data from their photometric
observations of M5, and Andy Ruggiero for sending us electronic files
containing the \citet{reid96} RRL light curves.  ACL, AS, and TvH
acknowledge support from NASA/STScI through grant number
HST-GO-08310-97A.  TvH gratefully acknowledges support by NASA through
LTSA grant 02-0045-0082. Guest User, Canadian Astronomy Data Centre,
which is operated by the Herzberg Institute of Astrophysics, National
Research Council of Canada.  The Digitized Sky Surveys were also used
in the course of this research.

%%%%%%%%%%%%%%%%%%%%%%%%%%%%%%%%%%%%%%%%%%%%%%%%%%%%%%%%%%%%%%%%%%%%%%%%

%%%%%%%%%%%%%%%%%%%%%%%%%%%%%%%%%%%%%%%%%%%%%%%%%%%%%%%%%%%%%%%%%%%%%%%%

\clearpage


%% FIGURE 1:  (IMAGES)
%\begin{figure}
%\plotone{f0a.gif}
%\caption{An image of the WF2 field in F814W after coadding all the
%long-exposure images and eliminating cosmic rays.  The image was used
%to detect and classify objects using SExtractor, and the resulting
%object list was used as input into ALLFRAME.    \label{fig_image_wf2} }
%\end{figure}

% FIGURE 1:  (2 CMDs)
%\figcaption{The calibrated, $VI$ CMD for M5. (a) All photometered
%objects are plotted, as well as the log(g) = 7.5 WD cooling track from
%\citet{bwb95} placed at a distance modulus
%of 14.70 mag.  (b) Only those objects are plotted that are uncrowded
%on the detector and likely to be stars, as indicated by the SExtractor
%class $>$ 0.75 (see Sec. \ref{sec_cmd}). \label{fig_cmd} }

\clearpage 

%% FIG 2: (WD FITS)
%\figcaption{The WD region of the M5 CMD showing the WDs used
%(open circles) and not used (bare error bars) in the distance fits.
%The calibrating field WDs, shifted to the fitted distance modulus of
%M5, are shown as triangles.  Panel (a) shows the results of the
%first-pass fit, while (b) is for the reselected WD candidates used in
%the second pass distance determination.
%\label{fig_wdfit} }

\clearpage 

%% FIG 3: (MS FIT)
%\figcaption{The weighted least squares fit of the M5 fiducial main
%sequence (solid line) to the field subdwarfs with {\it Hipparcos}
%parallaxes (open circles).  The adopted reddening and metallicity are
%indicated, along with the derived distance modulus.  \label{fig_sdfit}}

\clearpage 

%% FIG 4: (RRL BVIK)
%\figcaption{Apparent magnitudes of RRL in M5 compiled from the
%literature, plotted as a function of pulsation period, for the $B$,
%$V$, $I$, and $K$ passbands.  In each figure, the arrow indicates the
%mean apparent magnitude of the RRL. Squares and triangles indicate
%RRab and RRc pulsators, respectively.  Solid symbols indicate stars
%with light curves having complete phase coverage, while open symbols
%indicate stars with small phase gaps.  The crosses in (c) and (d)
%indicate the fundamentalized positions of the RRc stars, while the
%curves are the best fits represented by the coefficients in lines 1
%and 2 of Table \ref{tab_wasen}.  \label{fig_rrlmag} }

\clearpage 

% FIG 5: (WASENHEIT)
%\figcaption{Wasenheit functions for the M5 RRL as a function of
%fundamentalized period.  The symbols are as in Figure
%\ref{fig_rrlmag}.  Solid lines represent our least-squares fits to the
%data, while dashed lines indicate the predictions of models by 
%\citet{cassisi04}.  Error bars indicate systematic uncertainty in the fit
%due to uncertainties in the distance modulus and photometric
%zeropoints.  \label{fig_wasen} }

%%%%%%%%%%%%%%%%%%%%%%%%%%%%% TABLES %%%%%%%%%%%%%%%%%%%%%%%%%%%%%%%%%%%%%%%

%% UNDO THIS FOR FINAL FORMAT
%\clearpage

%%%%% TABLE 1: VISITS
\begin{deluxetable}{ccccc}
\tabletypesize{\scriptsize} 
%\tablewidth{50pc}
\tablecaption{Observation Log. \label{tab_visits}} 
\tablewidth{0pt} 
\tablehead{
\colhead{Visit} & 
\colhead{Date}  & 
\colhead{$N_{orbit}$} &
\colhead{Filter} & 
\colhead{Exposures}
} 
\startdata 
1 & 1999 Jul 25 & 1   &   F814W  & $2 \times 700$ sec, 600 sec, 60 sec, 5 sec \\
  &             & 2   &   F555W  & $3 \times 700$ sec, 60 sec, 5 sec \\
  &             & 3   &   F814W  & $3 \times 800$ sec \\
  &             & 4   &   F555W  & $3 \times 800$ sec \\
2 & 1999 Aug 9  & 1   &   F814W  & $2 \times 700$ sec, 600 sec, 60 sec, 5 sec \\
  &             & 2   &   F555W  & $3 \times 700$ sec, 60 sec, 5 sec \\
  &             & 3   &   F814W  & $3 \times 800$ sec \\
  &             & 4   &   F555W  & $3 \times 800$ sec \\
3 & 1999 Jul 29 & 1   &   F814W  & $2 \times 700$ sec, 600 sec, 60 sec, 5 sec \\
  &             & 2   &   F555W  & $3 \times 700$ sec, 60 sec, 5 sec \\
  &             & 3   &   F814W  & $3 \times 800$ sec \\
  &             & 4   &   F555W  & $3 \times 800$ sec \\
4 & 1999 Jul 17 & 1   &   F814W  & $2 \times 700$ sec, 600 sec, 60 sec, 5 sec \\
  &             & 2   &   F814W  & $3 \times 800$ sec \\
  &             & 3   &   F814W  & $3 \times 800$ sec \\
\enddata

%% Text for table notes should follow after the \enddata but before
%% the \end{deluxetable}. Make sure there is at least one \tablenotemark
%% in the table for each \tablenotetext.

%\tablenotetext{a}{Sample footnote for table~\ref{tbl-1} that was generated
%with the deluxetable environment}
%\tablenotetext{b}{Another sample footnote for table~\ref{tbl-1}}

%\tablecomments{The complete version of this table is in the electronic
%edition of the Journal.  The printed edition contains only a sample.}

\end{deluxetable}

%%%%%%%%%%%%%%%%%%%%%%%%%%%%%%%%%%%%%%%%%%%%%%%%%%%%%%%%%%%%%%%%%%%%%

%% UNDO THIS FOR FINAL FORMAT
%\clearpage

%%%%% TABLE 2: WF PHOTOMETRY
\begin{deluxetable}{ccccccccccc}
\tabletypesize{\scriptsize} 
%\tablewidth{50pc}
\tablecaption{WFPC2 Photometry. \label{tab_phot}} 
\tablewidth{0pt} 
\tablehead{
\colhead{ID} & 
\colhead{WF} & 
\colhead{$X$} & 
\colhead{$Y$} &
\colhead{$V$} & 
\colhead{$\epsilon_V$} &
\colhead{$I$} & 
\colhead{$\epsilon_I$} &
\colhead{ellip} &
\colhead{fwhm} &
\colhead{class} 
} 
\startdata 
 1 & 2 & 179.08 &  26.53 & 24.885 & 0.046 & 23.162 & 0.027 & 0.518 &
6.60 & 0.62 \\
 3 & 2 & 244.22 &  27.58 & 24.873 & 0.057 & 22.177 & 0.032 & 0.060 &
2.71 & 0.03 \\
 4 & 2 & 280.22 &  27.98 & 22.112 & 0.025 & 20.858 & 0.023 & 0.340 &
1.62 & 0.98 \\
%\medskip
 5 & 2 & 115.19 &  28.21 & 22.939 & 0.019 & 21.517 & 0.015 & 0.389 &
2.18 & 0.93 \\
... & ... & ... & ... & ... & ... & ... & ... & ... & ... & ... \\
 1 & 3 &  67.80 &  47.25 & 25.248 & 0.067 & 23.444 & 0.031 & 0.348 &
0.77 & 1.00 \\
 2 & 3 & 110.09 &  48.82 & 23.441 & 0.023 & 21.966 & 0.018 & 0.436 &
1.39 & 0.93 \\
 3 & 3 & 128.82 &  48.98 & 25.153 & 0.049 & 23.410 & 0.028 & 0.039 &
1.07 & 0.99 \\
%\medskip
 4 & 3 & 554.64 &  49.04 & 24.620 & 0.061 & 22.939 & 0.031 & 0.270
&12.37 & 0.04 \\
% & & & & & & & & & & \\
... & ... & ... & ... & ... & ... & ... & ... & ... & ... & ... \\
 1 & 4 & 768.48 &  42.40 & 24.820 & 0.035 & 23.039 & 0.022 & 0.183 &
2.06 & 0.16 \\
 2 & 4 & 674.81 &  42.46 & 24.657 & 0.042 & 22.892 & 0.028 & 0.201 &
2.88 & 0.02 \\
 3 & 4 & 599.91 &  43.10 & 23.487 & 0.021 & 21.951 & 0.021 & 0.186 &
2.27 & 0.58 \\
 4 & 4 & 561.39 &  43.15 & 21.601 & 0.019 & 21.170 & 0.016 & 0.133 &
1.56 & 0.99 \\
\enddata

%% Text for table notes should follow after the \enddata but before
%% the \end{deluxetable}. Make sure there is at least one \tablenotemark
%% in the table for each \tablenotetext.

%\tablenotetext{a}{Table entries with values of 9.999 or 9.99 and
%$N_{\rm pass} = 0$ indicate SExtractor data was not available for this
%object.}
%\tablenotetext{b}{Another sample footnote for table~\ref{tbl-1}}

\tablecomments{The complete version of this table is in the electronic
edition of the Journal.  The printed edition contains only a sample.}

\end{deluxetable}

%%%%%%%%%%%%%%%%%%%%%%%%%%%%%%%%%%%%%%%%%%%%%%%%%%%%%%%%%%%%%%%%%%%%%

%% UNDO THIS FOR FINAL FORMAT
%\clearpage

%%%% TABLE 3: PHOTOMETRY of FIELD WDs
\begin{deluxetable}{lrrrr}
\tabletypesize{\scriptsize} 
%\tablewidth{50pc}
\tablecaption{Photometry of Field White Dwarfs. \label{tab_wdphot}} 
\tablewidth{0pt} 
\tablehead{
\colhead{Star} & 
\colhead{$V$} & 
\colhead{$\epsilon_V$} &
\colhead{$V-I$} & 
\colhead{$\epsilon_{V-I}$}  
} 
\startdata 
WD~0644+375  &  9.96 & 0.08 & --0.23 & 0.11  \\
WD~1327--083 & 10.56 & 0.11 & --0.13 & 0.15  \\
WD~1935+327  & 11.06 & 0.11 &   0.00 & 0.16  \\
WD~2126+734  & 10.29 & 0.17 & --0.13 & 0.25  \\
WD~2326+049  & 11.24 & 0.15 &   0.00 & 0.22  \\
WD~2341+322  & 11.02 & 0.08 & --0.03 & 0.12  \\
\enddata

%% Text for table notes should follow after the \enddata but before
%% the \end{deluxetable}. Make sure there is at least one \tablenotemark
%% in the table for each \tablenotetext.

%\tablenotetext{a}{Sample footnote for table~\ref{tbl-1} that was generated
%with the deluxetable environment}
%\tablenotetext{b}{Another sample footnote for table~\ref{tbl-1}}

%\tablecomments{The complete version of this table is in the electronic
%edition of the Journal.  The printed edition contains only a sample.}

\end{deluxetable}

%%%%%%%%%%%%%%%%%%%%%%%%%%%%%%%%%%%%%%%%%%%%%%%%%%%%%%%%%%%%%%%%%%%%%

%% UNDO THIS FOR FINAL FORMAT
%\clearpage

% TABLE 4: Statistics on RRL apparent magnitudes.
 
%\makeatletter
%\def\jnl@aj{ApJ}
%\ifx\revtex@jnl\jnl@aj\let\tablebreak=\nl\fi
%\makeatother
 
\begin{deluxetable}{lrrrrrr}
\tabletypesize{\scriptsize} 
%\tablewidth{60pc}
\tablecaption{M5 RR Lyrae Magnitudes. \label{tab_rrlmag} }
\tablewidth{0pt} 
\tablehead{
 \colhead{Sample}       & 
 \colhead{$N_{RR}$}     &
 \colhead{$m(RR)$}      & 
 \colhead{sem}          &
 \colhead{$\sigma$}     &
 \colhead{median}       &
 \colhead{$M(RR)$}  
}
\startdata
$V$-band, all  & 79 &  15.024 &  0.009 &  0.082  & 15.05 & $+0.42 \pm 0.10$ \\
$V$-band, RRab & 53 &  15.025 &  0.011 &  0.082  & 15.05 & $+0.42 \pm 0.10$ \\
\medskip
$V$-band, RRc  & 26 &  15.023 &  0.017 &  0.084  & 15.05 & $+0.42 \pm 0.10$ \\
$B$-band, all  & 28 &  15.350 &  0.018 &  0.095  & 15.36 & $+0.71 \pm 0.10$ \\
$B$-band, RRab & 14 &  15.385 &  0.027 &  0.102  & 15.43 & $+0.74 \pm 0.10$ \\ 
\medskip
$B$-band, RRc  & 14 &  15.314 &  0.020 &  0.076  & 15.32 & $+0.67 \pm 0.10$ \\
$I$-band, all  & 37 &  14.562 &  0.017 &  0.104  & 14.58 & $+0.00 \pm 0.10$ \\
$I$-band, RRab & 24 &  14.520 &  0.018 &  0.087  & 14.52 & $-0.04 \pm 0.10$ \\ 
\medskip
$I$-band, RRc  & 13 &  14.641 &  0.024 &  0.086  & 14.63 & $+0.08 \pm 0.10$ \\
$K$-band, all  & 28 &  13.917 &  0.036 &  0.191  & 13.92 & $-0.59 \pm 0.12$ \\
$K$-band, RRab & 23 &  13.880 &  0.039 &  0.188  & 13.90 & $-0.63 \pm 0.12$ \\
$K$-band, RRc  &  5 &  14.086 &  0.045 &  0.100  & 14.08 & $-0.42 \pm 0.12$ \\
\enddata

\end{deluxetable}

%%%%%%%%%%%%%%%%%%%%%%%%%%%%%%%%%%%%%%%%%%%%%%%%%%%%%%%%%%%%%%%%%%%%%

%% UNDO THIS FOR FINAL FORMAT
%\clearpage

% TABLE 5: Wasenheit functions
 
%\makeatletter
%\def\jnl@aj{ApJ}
%\ifx\revtex@jnl\jnl@aj\let\tablebreak=\nl\fi
%\makeatother
 
\begin{deluxetable}{ccccrrc}
\tabletypesize{\scriptsize} 
%\tablewidth{30pc}
\tablecaption{Results of Period-Magnitude Fits. \label{tab_wasen}}
\tablewidth{0pt} 
\tablehead{
 \colhead{Line}      &
 \colhead{Function}  & 
 \colhead{$a$}       & 
 \colhead{$b$}       &
 \colhead{rms}       &
 \colhead{$N_{RR}$}  &
 \colhead{$R$}       
}
\startdata
1 & $I(RR)$   & $14.56 \pm 0.01$  & $-1.40 \pm 0.10$  & 0.05 & 36 & -- \\
2 & $K(RR)$   & $14.01 \pm 0.02$  & $-2.23 \pm 0.05$  & 0.04 &  4 & -- \\
3 & $M_I(RR)$ &  $0.00 \pm 0.10$  & $-1.40 \pm 0.10$  & 0.05 & 36 & -- \\
4 & $M_K(RR)$ & $-0.50 \pm 0.12$  & $-2.23 \pm 0.05$  & 0.04 &  4 & -- \\
5 & $W_0(BV)$ & $-0.48 \pm 0.10$  & $-2.94 \pm 0.24$  & 0.10 & 28 & 3.10 \\
6 & $W_0(VI)$ & $-0.58 \pm 0.10$  & $-2.68 \pm 0.15$  & 0.07 & 36 & 2.39 \\
7 & $W_0(VI)$ & $-0.65 \pm 0.10$  & $-2.82 \pm 0.16$  & 0.07 & 36 & 2.54\tablenotemark{a} \\
8 & $W_0(VK)$ & $-0.69 \pm 0.10$  & $-3.85 \pm 0.49$  & 0.14 & 20 & 0.13 \\
\enddata
%5 & $W_0(BV)$ & $-0.477 \pm 0.020$ & $-2.94 \pm 0.24$ & 0.10 & 28 & 3.10 \\
%6 & $W_0(VI)$ & $-0.576 \pm 0.011$ & $-2.68 \pm 0.15$ & 0.07 & 36 & 2.39 \\
%7 & $W_0(VI)$ & $-0.646 \pm 0.012$ & $-2.82 \pm 0.16$ & 0.07 & 36 & 2.54\tablenotemark{a} \\
%8 & $W_0(VK)$ & $-0.688 \pm 0.031$ & $-3.85 \pm 0.49$ & 0.14 & 20 & 0.13 \\
%\enddata

\tablenotetext{a}{For comparison with models of \citet{cassisi04}.}

\end{deluxetable}

%% The following command ends your manuscript. LaTeX will ignore any text

\end{document}